\documentclass[conference]{IEEEtran}
\usepackage{cite}
\usepackage{graphicx}
\usepackage{amsmath,amssymb}
\usepackage{url}
\usepackage{hyperref}
\usepackage{subcaption}
\usepackage{placeins}
\usepackage{float}

\begin{document}

\title{SARCLIP: A Scalable CLIP-Based Retrieval System for Seventeenth-Century Spanish American Notary Records}

\author{
\IEEEauthorblockN{Chandrasekhar Syamala\IEEEauthorrefmark{1}\IEEEauthorrefmark{5},
Parshad Suthar\IEEEauthorrefmark{2}\IEEEauthorrefmark{5},
Hulayyil Alshammari\IEEEauthorrefmark{2},
Viviana Grieco\IEEEauthorrefmark{3},
Praveen Rao\IEEEauthorrefmark{2}}
\IEEEauthorblockA{\IEEEauthorrefmark{1}MU School of Medicine, University of Missouri, Columbia, MO, USA}
\IEEEauthorblockA{\IEEEauthorrefmark{2}Dept. of Electrical Engineering and Computer Science, University of Missouri, Columbia, MO, USA}
\IEEEauthorblockA{\IEEEauthorrefmark{3}University of Missouri--Kansas City, Kansas City, MO, USA}
\IEEEauthorblockA{\IEEEauthorrefmark{5}Equal contribution}
}

\maketitle

\begin{abstract}
Historical manuscript archives resist standard text search due to inconsistent handwriting, 
archaic orthography, and the absence of reliable transcriptions at scale. We present SARCLIP (Spanish American Notary Records Meets CLIP), a deployed retrieval system for the National Archives of Argentina's seventeenth-century Spanish American notary records, a corpus of more than 13.6 million word-image patches spanning 100 of 102 microfilm rolls (``rollos''). SARCLIP is built on a CLIP ViT-B/16 model contrastively fine-tuned on paleography-expert-annotated data, and extends prior work by (1) scaling approximate nearest-neighbor retrieval to the near-complete corpus via a FAISS index, (2) refining top-$k$ results through pseudo-relevance feedback (Rocchio), and (3) closing a human-in-the-loop cycle through visual document browsing, canvas-based patch annotation, and periodic model retraining. Unlike the system's initial research prototype, which evaluated retrieval on a small five-rollo subset, SARCLIP is demonstrated as a complete, interactive tool operating over the near-complete corpus. Attendees experience the full search, browse, annotate, and retrain workflow live during this demonstration.
\end{abstract}


\section{Introduction}

Archives of colonial-era notarial records (wills, powers of attorney, property transfers, debt instruments) are of substantial value to historians, but their handwritten, multi-scribe, multi-decade nature makes them resistant to standard full-text search \cite{alrasheed2021}. Notarial scribes working across a century and a half produced records with inconsistent letterforms, abbreviation conventions, and archaic orthography that vary not only scribe to scribe but often within a single scribe's career, making transcription-first approaches brittle at corpus scale. The SARCLIP corpus digitizes 102 rollos (microfilm rolls) from the National Archives of Argentina, comprising an estimated 60{,}000 document pages and, in the current index, 13{,}619{,}093 segmented word-level patches.

Prior published work on this corpus, KGSAR \cite{prasanna2022}, combined domain-retrained OCR with a knowledge graph for document retrieval. As the first stage of the present work, we fine-tuned a Contrastive Language-Image Pretraining (CLIP) \cite{radford2021} model on approximately 40{,}000 hand-annotated word patches from two rollos, retrieving via direct embedding similarity search over a small evaluation subset. This substantially outperformed the KGSAR pipeline, raising mean average precision (mAP) from 0.267 to 0.809 in an expert-verified evaluation over held-out rollos.

This demo presents SARCLIP's evolution from a research prototype, evaluated on five rollos, into a deployed, interactive tool operating over the near-complete 102-rollo corpus. We highlight three developments that distinguish the deployed system from these earlier stages:

\begin{itemize}
\item \textbf{Scale.} The system indexes 100 of 102 rollos (13.6M patches) via an approximate nearest-neighbor index, rather than a small evaluation subset.
\item \textbf{Retrieval refinement.} A pseudo-relevance feedback (PRF) pass sharpens results beyond single-shot embedding similarity.
\item \textbf{A complete interactive workflow.} Search is now one of four integrated capabilities (search, visual browsing, annotation, and periodic retraining) forming a closed human-in-the-loop system rather than a static retrieval demo.
\end{itemize}

The result is a tool historians and paleographers can use directly, closing the loop from query to correction to retraining within a single interface.
\section{Related Work}

Conventional OCR systems face challenges with irregular historical scripts; previous assessments on this dataset revealed that standard OCR tools (Tesseract, Kraken, Calamari-OCR) achieved zero correct words on a 670-word annotated test set, whereas models retrained for the domain performed better but still needed considerable retraining effort \cite{alrasheed2021}. The KGSAR system \cite{prasanna2022} integrated retrained OCR with a Knowledge Graph for organized retrieval, yet its effectiveness was closely linked to the particular handwriting styles found in its training dataset.

As the first stage of the present work, we fine-tuned CLIP on word-image patches directly, avoiding an intermediate OCR transcription phase and significantly surpassing the KGSAR
baseline (Section~\ref{sec:eval}). That initial assessment, however, was performed on a five-rollo subset using a single-pass, non-scaled similarity search and offered only a basic search page without browsing, annotation, or retraining capabilities. SARCLIP advances this area of research into a practical application: it enhances retrieval across the entire corpus, introduces a second-pass re-ranking phase, and builds visual browsing, annotation, and
retraining directly into the tool rather than leaving them as separate research efforts.

The OCR benchmark~\cite{alrasheed2021} and KGSAR~\cite{prasanna2022} discussed above are part of a broader NEH-supported research program on this archive. Companion work in this program evaluated deep-learning techniques for content extraction from these records~\cite{alrasheed2021sumac}, explored few-shot learning to recognize handwritten words from minimal labeled samples~\cite{alrasheed2023fewshot}, and used the corpus to
fine-tune Spanish large language models on colonial-era text~\cite{sarker2024jcdl}; KGSAR itself was presented as a demonstration at ISWC 2022. Each of these efforts addressed a component capability; whether character recognition, content
extraction, low-label word recognition, or language modeling, and each was evaluated offline on limited subsets of the archive. The proposed demonstration differs in both scope and form: SARCLIP unifies retrieval over the near-complete 13.6M-patch corpus with PRF re-ranking, visual browsing, in-interface annotation, and human-in-the-loop retraining in one deployed, interactive
tool. None of the earlier systems combined these in a single interface.

CLIP-like contrastive image-text models have been applied to cross-modal retrieval well beyond their original web-image training data, which motivates applying one here without a preliminary
handwritten text recognition (HTR) stage. Pseudo-relevance feedback is a standard text-retrieval technique that refines the top-$k$ results using signals from an initial retrieval pass\cite{rocchio1971}. We adapt it to a purely visual embedding space: instead of rewriting a bag-of-words query, we merge the original query embedding with the average embedding of the top-ranked results.
\section{System Architecture}

\begin{figure}[t]
\centering
\includegraphics[width=\columnwidth]{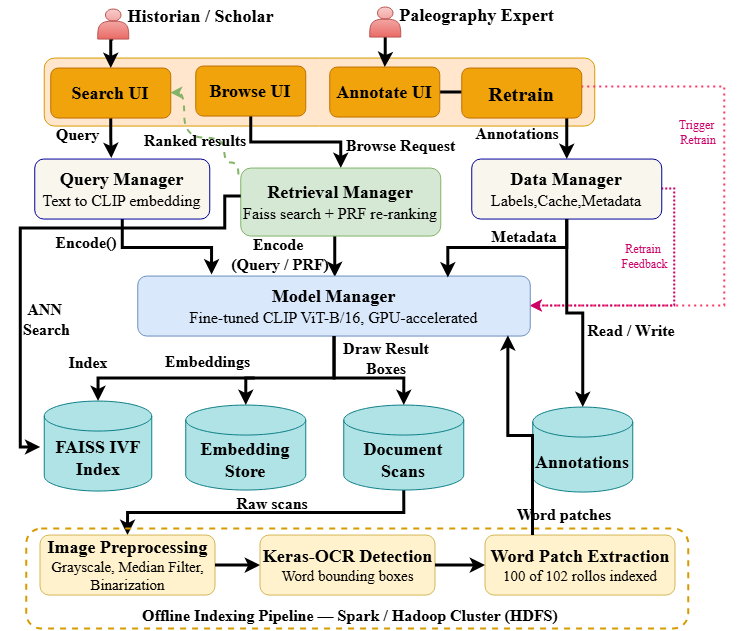}
\caption{SARCLIP system architecture. Online (top): a query flows
through CLIP text encoding, direct FAISS search, and PRF re-ranking to
produce ranked results, with a parallel visual browse path. Offline
(dashed container, bottom): document scans are preprocessed, word
regions are detected with Keras-OCR, and patches are extracted on the
Spark/Hadoop cluster before CLIP encoding populates the embedding store
and FAISS index. Expert annotations periodically retrain the CLIP
checkpoint, closing the loop (dashed).}
\label{fig:architecture}
\end{figure}

Fig.~\ref{fig:architecture} illustrates the SARCLIP pipeline. A text
query is processed using the fine-tuned CLIP text encoder and then
compared to a FAISS index \cite{johnson2019faiss} containing
13.6 million word-patch embeddings (IVF, \texttt{nprobe}=128; an HNSW
index \cite{malkov2020hnsw} is being assessed as a potential
higher-recall option at this scale). The leading candidates undergo a
pseudo-relevance feedback (PRF) \cite{rocchio1971} process: the top-3
results are re-embedded, and a combined query vector
($0.7\times$ text embedding $+ 0.3\times$ average image embedding)
facilitates a subsequent retrieval pass, which is then merged and
deduplicated with the initial results.

\subsection{Distributed Corpus Construction}

The FAISS index and embedding store are built offline, separately from
the online query path shown in Fig.~\ref{fig:architecture}. Building on
the distributed infrastructure used to scale the original KGSAR
pipeline, word-patch extraction runs on an eight-node CloudLab cluster
\cite{duplyakin2019cloudlab} using Apache Spark
\cite{zaharia2010spark} over the Hadoop Distributed File System (HDFS)
\cite{white2009hadoop}. Document scans are loaded into HDFS,
parallelized into Spark Resilient Distributed Datasets (RDDs), and
partitioned across nodes for preprocessing (grayscale conversion,
denoising, and binarization) and word-region bounding-box detection
with Keras-OCR, producing the segmented word patches later indexed by
FAISS.

Distributing this pipeline across the cluster reduced end-to-end
processing time for the corpus's approximately 60{,}000 document pages
by roughly an order of magnitude compared with single-machine
processing. Once patches are extracted, the fine-tuned CLIP model
described below encodes each patch to populate the embedding store shown
in Fig.~\ref{fig:architecture}; this encoding step runs on the GPU
server rather than the Spark cluster. Thus, the Model Manager box in
Fig.~\ref{fig:architecture} represents CLIP fine-tuning and embedding
generation specifically, rather than the upstream distributed
extraction stage.

\subsection{Fine-Tuned Retrieval Model}

In the first stage of this work, we fine-tuned CLIP ViT-B/16
\cite{dosovitskiy2021vit} through contrastive learning
(Adam, learning rate $1{\times}10^{-7}$, weight decay
$1{\times}10^{-4}$, batch size 32, 500 epochs) using
paleography-expert-annotated word patches from Rollos 38 and 40.
We use this domain-adapted model as SARCLIP's retrieval model; fine-tuning gives it greater robustness to variations in historical handwriting.

\subsection{Retrieval Quality Evaluation}
\label{sec:eval}

We evaluated the fine-tuned model against the KGSAR baseline
under identical conditions. Both systems were queried through
their respective web interfaces with the same 75 keywords
selected by paleography experts, over approximately 3,000
documents from five rollos (39, 41, 42, 43, and 44). These
rollos were fully held out: the model was fine-tuned only on
Rollos 38 and 40, so no test rollo appeared in training, so the evaluation measures generalization across rollos rather than within-rollo memorization.

For each keyword, experts reviewed the top-5 retrieved word
patches and marked whether each patch matched the query
term. Average precision was computed per keyword and
averaged across all 75 queries to obtain mAP. As shown
in Table~\ref{tab:retrieval-quality}, the fine-tuned CLIP model
reached an mAP of 0.809 compared with 0.267 for KGSAR,
roughly a threefold improvement. The gain comes from
matching query embeddings directly against word-image
embeddings in a shared representation space, which avoids
the OCR transcription errors that limited the keyword-lookup
approach in KGSAR. Errors such as mis-transcribed
abbreviations and inconsistent letterforms caused KGSAR to
miss valid matches entirely, while the embedding-based
approach retrieves them based on visual similarity alone.This evaluation predates the full-corpus index; we do not claim these numbers hold at 13.6M patches, where crowding effects are expected to reduce precision@k.

\begin{table}[t]
\centering
\caption{Retrieval quality: fine-tuned CLIP vs. prior KGSAR
baseline, mAP over 75 expert-selected keywords on five
held-out rollos.}
\label{tab:retrieval-quality}
\begin{tabular}{lc}
\hline
\textbf{Method} & \textbf{mAP} \\
\hline
KGSAR (knowledge-graph baseline)~\cite{prasanna2022} & 0.267 \\
Fine-tuned CLIP (this work) & \textbf{0.809} \\
\hline
\end{tabular}
\end{table}

\subsection{Query and Data Management}

The Query Manager processes incoming text queries using the refined
CLIP text tower and sends their embeddings to the Retrieval Manager,
which manages FAISS index search and PRF re-ranking. To separate query
representation from index-specific retrieval logic, the Query Manager
is kept apart from the Retrieval Manager, enabling independent
modification, optimization, or replacement of the CLIP query encoder
and the FAISS/PRF search strategy.

The Data Manager oversees ancillary state that is not included in the
index itself: browse-cluster caches, patch-to-page metadata, and labels
from user-submitted annotations. Isolating this state from the FAISS
index enables the Data Manager's contents to be updated, for example
when new annotations are submitted, without requiring immediate index
reconstruction; index regeneration is reserved for scheduled model and
corpus updates.

\subsection{Visual Browse}

Sampled patches are grouped with MiniBatch K-Means
\cite{sculley2010minibatch} ($k=6$) over their fine-tuned CLIP
embeddings, letting users browse the collection by visual similarity in
handwriting style, ink density, and layout. MiniBatch K-Means was chosen
for building browse groups from the large patch corpus because it
reduces the computational cost of centroid-based clustering on large
embedding collections compared with full batch K-Means.

To provide a small, navigable set of visual groups without
over-fragmenting the browsing space, the current prototype uses $k=6$
as an interface-level design parameter. It is therefore not meant to reflect the collection's ideal or intrinsic number of semantic
categories. We call these groups \emph{visual clusters}, rather than
semantic or thematic categories: the embedding space is trained for
image--text patch alignment and does not encode document-level topic
structure.

\subsection{Annotate and Retrain}

A canvas tool lets annotators zoom, pan, and draw bounding boxes around
word segments, building a supervised dataset that extends the initial
two-rollo training set. The resulting annotations are maintained by the
Data Manager and can be incorporated during scheduled fine-tuning of
the CLIP model, allowing expert feedback to improve subsequent model
checkpoints without coupling annotation activity directly to the live
retrieval index.

\subsection{Deployment}

The system operates as a Flask backend integrated with a ReactJS
frontend, utilizing CLIP inference and index search accelerated on a
single NVIDIA RTX 5000 Ada GPU (32 GB VRAM). All four functionalities
(search, browse, annotate, and retrain) are accessible through a single
web interface, allowing users to navigate among them without changing
context or using separate tools.
\section{Demo Description}
Attendees will interact with the live SARCLIP web interface across four scenarios:

\begin{enumerate}
\item \textbf{Search.} Participants input Spanish notarial keywords (e.g., \textit{poder}, \textit{testigo}, \textit{escribano}) and see ranked word-patch results displayed with bounding-box highlights on the source document page, fetched in real time from the near-complete 13.6M-patch, $\sim$100-rollo index.
\item \textbf{Browse.} Participants navigate the collection through visually grouped word patches, each featuring representative thumbnails, allowing them to transition from a cluster directly into a keyword search.
\item \textbf{Annotate.} Participants choose a retrieved patch, magnify its source page, and create a bounding box to mark a word, illustrating how new supervision is gathered directly through user engagement.
\item \textbf{Retrain.} Participants initiate a retraining process using newly acquired labels and monitor the revised label count and training progress, showcasing the complete human-in-the-loop cycle.
\end{enumerate}

The scenarios are meant to be executed in order, allowing one attendee session to complete the entire cycle: a search surfaces a result the model ranks with low similarity, the attendee correctly annotates it, and a following retraining session shows the model taking that correction into account, so attendees can see the human-in-the-loop design in Fig.~\ref{fig:architecture} run end to end.

This submission includes a brief recorded walkthrough of all four scenarios, and the live system will be accessible for hands-on experience during the conference demonstration session.

\begin{figure}[H]
\centering
\begin{subfigure}[b]{\linewidth}
\centering
\includegraphics[width=0.8\linewidth]{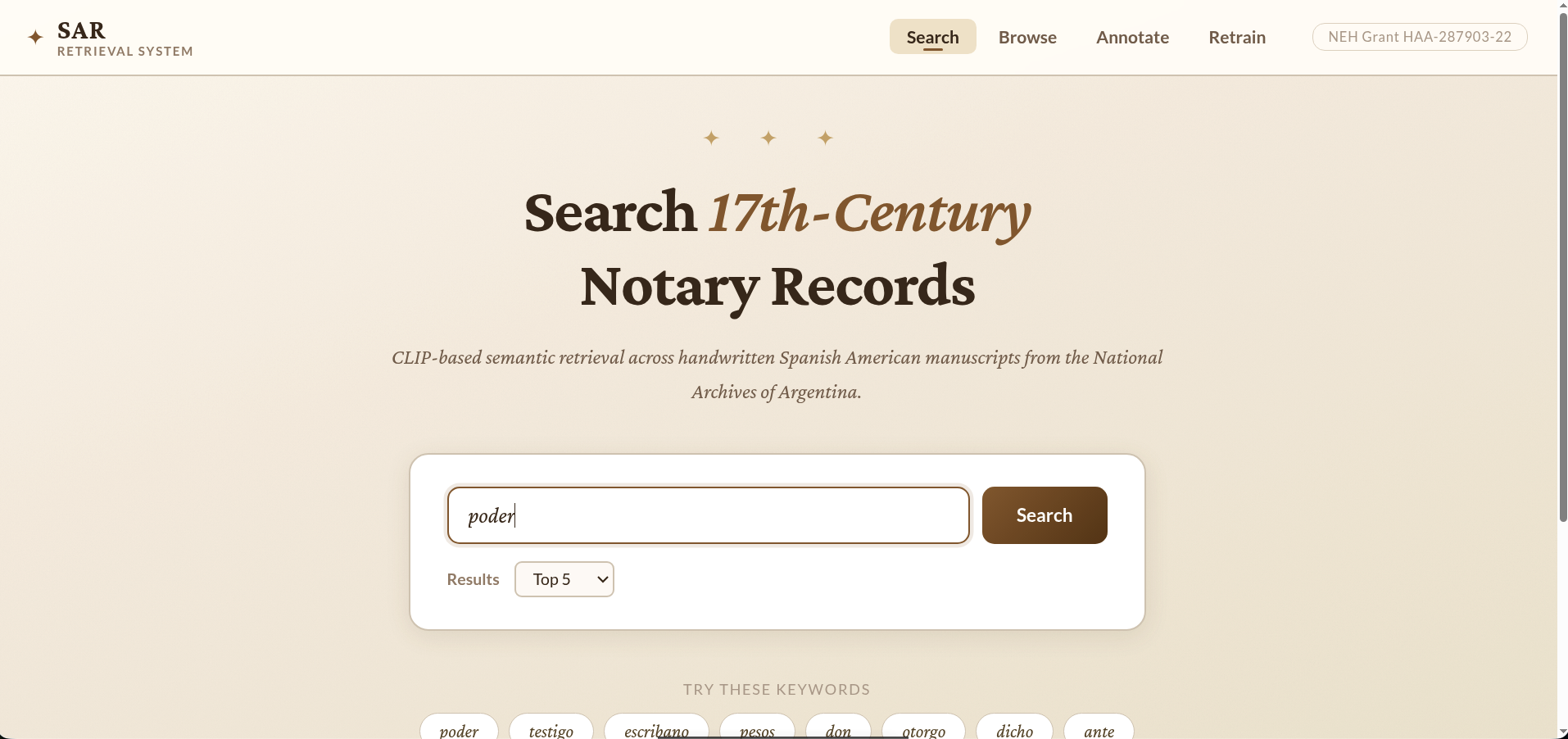}
\caption{Query entry}
\label{fig:ui-query}
\end{subfigure}
\begin{subfigure}[b]{\linewidth}
\centering
\includegraphics[width=0.8\linewidth]{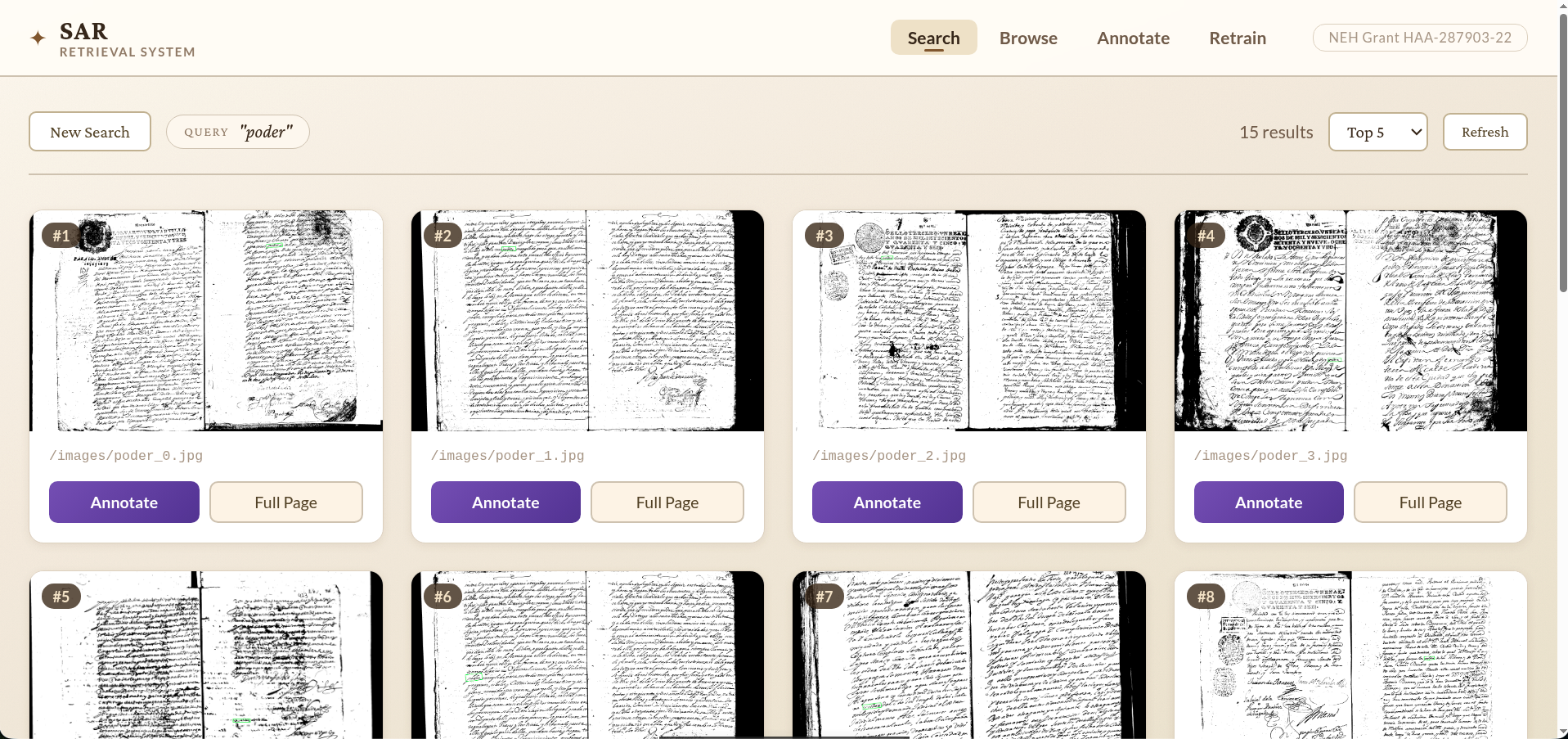}
\caption{Ranked results}
\label{fig:ui-results}
\end{subfigure}
\begin{subfigure}[b]{\linewidth}
\centering
\includegraphics[width=0.8\linewidth]{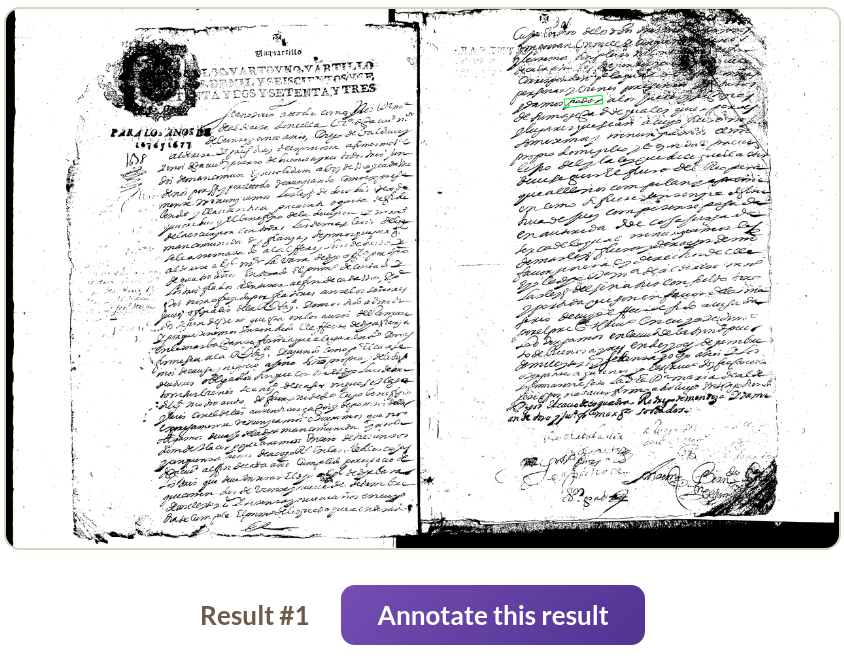}
\caption{Retrieved patch, boxed}
\label{fig:ui-zoomed}
\end{subfigure}
\caption{SARCLIP search interface for query \textit{poder}: (a) the query entry screen, (b) ranked word-patch results returned from the full corpus, (c) a matched word patch with bounding-box highlight.}
\label{fig:ui-screenshot}
\end{figure}

\FloatBarrier
\section{Conclusion}

SARCLIP takes a fine-tuned CLIP retrieval model from a five-rollo research evaluation to a deployed tool that runs over the near-complete 13.6M-patch collection of seventeenth-century Spanish
American notarial documents. Retrieval is refined with pseudo-relevance feedback, and search, visual browsing, annotation, and retraining are combined in one interface.

We see three directions for future work. First, we plan to expand fine-tuning by folding in labels gathered through the annotation tool as attendees and scholars use it, growing the training set
beyond the initial two rollos. Second, we will compare HNSW against IVF indexing over the full corpus to weigh recall against latency as the index grows toward all 102 rollos. Third, we intend
to investigate handwritten text recognition (HTR) for this script, which would let retrieval draw on both visual and textual evidence rather than visual embeddings alone.
\section*{Acknowledgment}
This work was supported by the National Endowment for the Humanities (NEH), Award No.\ HAA-287903-22.

\bibliographystyle{IEEEtran}
\bibliography{references}

\end{document}